\documentclass[journal]{IEEEtran}

\usepackage{cite}

\ifCLASSINFOpdf
  \usepackage[pdftex]{graphicx}
\else
  \usepackage[dvips]{graphicx}
\fi

\usepackage[cmex10]{amsmath}
\usepackage{algorithmic}

\usepackage{array}

\ifCLASSOPTIONcompsoc
  \usepackage[caption=false,font=normalsize,labelfont=sf,textfont=sf]{subfig}
\else
  \usepackage[caption=false,font=footnotesize]{subfig}
\fi

\usepackage{fixltx2e}
\usepackage{stfloats}

\usepackage{url}

\usepackage{bm} 
\usepackage{amssymb}
\usepackage{xcolor} 
\usepackage{booktabs} 
\usepackage{enumerate} 
\usepackage{svg} 
\usepackage{multirow} 

\begin{document}

\title{A Deep Learning Approach for SAR Tomographic Imaging of Forested Areas}

\author{Zoé~Berenger,    Loïc~Denis,~\IEEEmembership{Senior~Member,~IEEE,}
        Florence~Tupin,~\IEEEmembership{Senior~Member,~IEEE,}
        Laurent~Ferro-Famil,~\IEEEmembership{Member,~IEEE,}
        and~Yue~Huang
\thanks{This project has been funded by the Futur \& Ruptures PhD program of the Fondation Mines-Telecom, and partially funded by ASTRAL project (ANR-21-ASTR-0011).}
\thanks{Z. Berenger and F. Tupin are with LTCI, Télécom Paris, Institut polytechnique de Paris, Paris, France (e-mail: name.surname@telecom-paris.fr).}
\thanks{L. Denis is with the Laboratoire Hubert Curien, UMR 5516, CNRS, Institut d'Optique Graduate School, Univ. Lyon, UJM-Saint-Étienne, Saint-Étienne 42023, France (e-mail: loic.denis@univ-st-etienne.fr).}
\thanks{L. Ferro-Famil is with ISAE-SUPAERO and CESBIO, University of Toulouse, Toulouse, France (e-mail: laurent.ferro-famil@isae-supaero.fr).}
\thanks{Y. Huang is with CESBIO, University of Toulouse, Toulouse, France (e-mail: yhuang228@gmail.com).}}

\markboth{Submitted to IEEE Geoscience and Remote Sensing Letters, January~2023}%
{Submitted to IEEE Geoscience and Remote Sensing Letters, January~2023}

\maketitle

\begin{abstract}
Synthetic aperture radar tomographic imaging reconstructs the three-dimensional reflectivity of a scene from a set of coherent acquisitions performed in an interferometric configuration. In forest areas, a large number of elements backscatter the radar signal within each resolution cell. To reconstruct the vertical reflectivity profile, state-of-the-art techniques perform a regularized inversion implemented in the form of iterative minimization algorithms. We show that light-weight neural networks can be trained to perform the tomographic inversion with a single feed-forward pass, leading to fast reconstructions that could better scale to the amount of data provided by the future BIOMASS mission. We train our encoder-decoder network using simulated data and validate our technique on real L-band and P-band data.
\end{abstract}


\IEEEpeerreviewmaketitle

\section{Introduction}

\IEEEPARstart{S}{ynthetic} aperture radar (SAR) tomography (TomoSAR) uses a 2D aperture to perform 3D imaging. In the case of a narrow-band radar waveform and under the widely adopted Born approximation at order 1 \cite{ferrofamil_principles_2016}, the imaging process simplifies to the 1D spectral analysis of a set of co-registered 2D SAR images \cite{Gini_2005_bf_capon}. It aims at reconstructing, for each 2D location, reflectivity profiles in the direction orthogonal to the radar line-of-sight.
As illustrated in \cite{huang_concealed_2017}, parametric spectral estimation approaches, which have been widely used for the characterization of urban areas, such as high-resolution techniques \cite{huang_concealed_2017} or \emph{Compressive Sensing (CS)}-inspired regularized least squares minimization \cite{Gilda_2011_CS, Rambour_2020_review}, fail to adequately reconstruct the response of forested environments. They indeed estimate a small set of discrete point-like scattering sources, instead of a continuous function, known to represent well the reflectivity of such volumetric media \cite{aghababaee_2020}.

Among the wide range of existing non-parametric spectral estimation techniques \cite{stoica_SpectralAnalysis}, the \emph{beamformer}, i.e. the discrete Fourier transform, and \emph{Capon}'s filter, also called the adaptive \emph{beamformer}, are the most widely used to perform TomoSAR focusing over forests. The \emph{beamformer} has a coarse resolution and creates sidelobes, whereas \emph{Capon}'s sidelobe reduction capability comes at the cost of radiometric accuracy \cite{stoica_SpectralAnalysis}. A parametric solution based on the use of a sparsifying basis able to approximate a continuous function using a small set of coefficients was proposed in \cite{Aguilera_2013_WV}. This approach, named \emph{wavelet-based CS}, used a \emph{CS}-inspired optimization to determine a reflectivity profile constructed using an orthogonal wavelet matrix and a regularized number of wavelet coefficients. Another approach, using a small number of parametric basis functions, was proposed in \cite{Laurent_2022_Dict}.
The regularized inversion of a linear model of the covariance matrix leads to much more accurate reconstructions, nevertheless the latter estimators, which are themselves non-linear, require costly iterative minimization algorithms that impede their application to large-scale datasets.

In many imaging domains, deep learning has made it possible to reduce computation time while maintaining high-resolution results. This potential has been harnessed for SAR tomography in \cite{Qian_2021_SARISTA}, where the authors unroll an Iterative Shrinkage Thresholding Algorithm (ISTA) to solve the $L_2 - L_1$ norm minimization problem posed by \emph{CS} in urban areas.
This approach has been improved in \cite{Qian_2022_gamma, Yunqiao_2022_Adatomo}, yet most deep learning techniques for SAR tomography over forests focus on ground and canopy height estimation using LiDAR data as a reference \cite{Yang_2022_DLheight}.

This paper presents a supervised deep learning method for tomographic SAR reconstruction in forested areas. Its objective is to recover the reflectivity profile from the coarse profile obtained by the \emph{beamforming} algorithm. We first use a physics-inspired generation model and interferometric baselines matching our SAR dataset to simulate reflectivity distributions and associated measurements. We then train a network with a light-weight architecture to learn a low-dimensional latent representation of these simulated profiles and to recover the original profiles free from \emph{beamforming} artifacts. Finally, the neural network is evaluated on real \emph{beamforming} profiles at L-band and P-band and compared to several methods, showing promising performances both in terms of reconstruction quality and computation time.

\section{Methodology}

\subsection{Problem formulation}
\label{section:pb}

The tomographic signal measured over $N$ SAR acquisitions, $\mathbf{y} \in \mathbb{C}^{N}$, may be formulated as the sum of the $N_s$ contributions originating from the considered 2D resolution cells as:
\begin{equation}
	\mathbf{y} = \sum \limits _{k=1}^{N_s} s_k \, \mathbf{a}(z_k) + \boldsymbol{\epsilon} = \mathbf{A} \mathbf{s} + \boldsymbol{\epsilon}
	\label{eq:model}
\end{equation}
where $\mathbf{s}=[s_1, \ldots, s_{N_s}]^T$ contains the complex reflection coefficient of the observed scatterers, $\mathbf{a}(z)=[1,\ldots,\text{e}^{j k_{z}^{(N)} z }]^T$ is a steering vector and models the interferometric phase with $k_{z}^{(n)} z$ corresponding to the phase seen on the $n$-th image for a scatterer located at height $z$, $\mathbf{A}=[\mathbf{a}(z_1), \ldots, \mathbf{a}(z_{N_s})]$ is the sensing matrix, while $\boldsymbol{\epsilon}$ stands for the system noise, distributed according to a circular complex Gaussian distribution of covariance $\sigma_{\epsilon}^2 \mathbf{I}_N$.

As mentioned earlier, over forested areas, the vertical density of reflectivity $\mathbf{s}$ is well modelled by a speckle-affected continuous function, and hence may be represented in (\ref{eq:model}) by a large number $N_s \gg N$ of uncorrelated source reflectivities distributed over a vector $\mathbf{z} \in \mathbb{R}^{N_z}$ of $N_z$ discrete heights. The covariance matrix of $\mathbf{s}$ is diagonal and is given by $\boldsymbol{\Sigma}_{\mathbf{ss}}= \mathbb{E}[\mathbf{s}\mathbf{s}^H] = \text{diag}(\mathbf{p})$, with $\cdot^H$ the Hermitian transpose operator. The measured signal covariance matrix can then be expressed as:
\begin{equation}
	\mathbf{\Sigma} = \mathbb{E}[\mathbf{y}\mathbf{y}^H] = \mathbf{A} \text{diag}(\mathbf{p}) \mathbf{A}^H + \sigma_{\epsilon}^2 \mathbf{I}_N
	\label{eq:cov_forest}
\end{equation}

In practice, this quantity is estimated using $L$ independent realizations of the measured vector, $\{\mathbf{y}_l\}_{l=1}^{L}$, sampled in the neighborhood of a 2D location. The corresponding sample covariance matrix is given by $\mathbf{\widehat{\Sigma}} = \frac{1}{L} \sum_{l=1}^{L} \mathbf{y}_l \mathbf{y}^H_l$.
The objective of forest TomoSAR imaging is the estimation, or at least the characterization,  of $\mathbf{p} \in \mathbb{R}^{N_z}_{+}$ from $\mathbf{\widehat{\Sigma}}$.
The sample correlation matrix, $\mathbf{\widehat{R}} = \text{diag}(\mathbf{q}) \boldsymbol{\widehat{\Sigma}} \text{diag}(\mathbf{q}) $, with $q_i = 1/\sqrt{\widehat{\boldsymbol \Sigma}_{ii}}$, represents a version of  $\mathbf{\widehat{\Sigma}}$, in which the intensity of each image is scaled to 1. Such a representation, independent of the observed absolute reflectivity, may also be used during specific steps of the 3D imaging process.

\subsection{From fixed dictionaries to learned representations: strategies for vertical profile reconstruction}

As shown in \cite{huang_concealed_2017, Aguilera_2013_WV}, forest reflectivity profiles $\mathbf{p}$ can typically be approximated as a linear combination of a few basis functions:
\begin{equation}
    \mathbf{p} \approx \boldsymbol{\Psi} \boldsymbol{\alpha}
    \label{eq:dictionary}
\end{equation}
where $\boldsymbol{\Psi} \in \mathbb{R}^{N_z \times N_{\alpha}}$ is the dictionary whose columns are the basis functions and $\boldsymbol{\alpha} \in \mathbb{R}^{N_{\alpha}}$ is the vector of weights. For an adequately chosen basis, only a few functions at a time are necessary to approximate a given profile. The vector $\boldsymbol{\alpha}$ is then sparse: most coefficients in $\boldsymbol{\alpha}$ are zero and the number of non-zeros $\|\boldsymbol{\alpha}\|_0$ is small, as is the L1 norm $\|\boldsymbol{\alpha}\|_1$ which is often used as a proxy to measure sparsity. 
The reflectivity profile $\widehat{\mathbf p}^{\text{(CS)}}=\boldsymbol{\Psi} \widehat{\boldsymbol{\alpha}}$ can be estimated using the following minimization problem \cite{Aguilera_2013_WV}:

\begin{align}
   \widehat{\mathbf p}^{\text{(CS)}}
   &=\mathop{\text{arg min}}_{\mathbf p\geq \mathbf 0}\; \|\mathbf{A} \text{diag}(\mathbf{p}) \mathbf{A}^H 
   -\widehat{\mathbf \Sigma}\|_{\text{F}}^2 + \lambda\|\boldsymbol{\Psi}^{\dagger} \mathbf{p}\|_1\,.
    \label{eq:minpbwavelets}
\end{align}
where $\lambda$ is a hyper-parameter responsible for balancing the weight of the sparsity constraint, with respect to the data-fidelity term defined by the Frobenius norm.
The pseudo inverse matrix, $\boldsymbol{\Psi}^{\dagger}$, in the L1 norm term verifies $\boldsymbol{\Psi} \boldsymbol{\Psi}^{\dagger} \mathbf{p}=\boldsymbol{\alpha}$, and may be replaced with $\boldsymbol{\Psi}^{H}$ in the case of a unitary orthogonal basis.
Solving this problem typically requires several tens or hundreds of iterations of a minimization algorithm and the proper tuning of the hyper-parameter $\lambda$.

\smallskip
Rather than using a linear model for the profiles $\mathbf p$, non-linear models such as the generative model discussed in paragraph \ref{sec:genemodel} can be used (e.g., Gaussian mixture models, or exponential profiles \cite{Laurent_2022_Dict}). However, inverting such models can be difficult due to identifiability issues and require additional constraints such as order constraints as recently proposed in a different modality of remote sensing for the estimation of phenological parameters from NDVI time series \cite{zerah2022physics}. Our early experiments have shown that it was preferable to jointly learn a non-linear model for $\mathbf p=m(\boldsymbol{\alpha})$ and an estimator $\widehat{\boldsymbol{\alpha}}$ by training a deep neural network with an encoder-decoder architecture. The code $\boldsymbol{\alpha}$ in the latent space learnt by the encoder is then turned into a profile by the decoder. The use of a simple feed-forward pass of a light-weight neural architecture leads to an extremely efficient tomographic inversion method, as described in the following paragraphs.

\subsection{Proposed method}

\begin{figure*}
    \centering
    \includegraphics[width=0.99\textwidth]{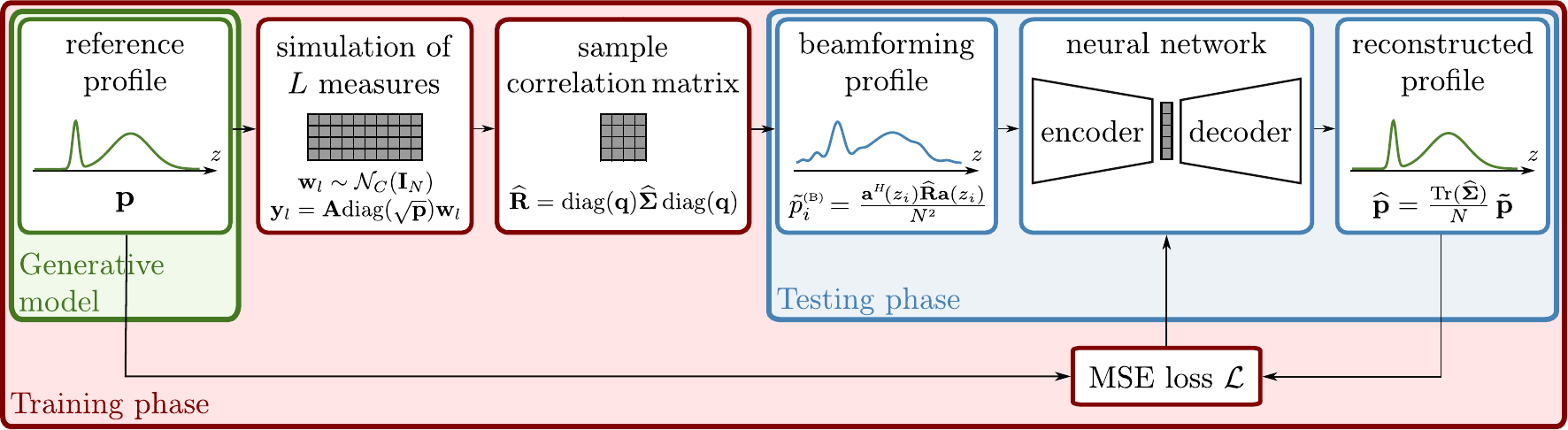}
    \caption{Pipeline of the data simulation, training and testing process of the proposed method.}
    \label{fig:pipeline}
\end{figure*}

The different steps of our method are presented in Fig. \ref{fig:pipeline}.
Starting from a ground truth profile created with a defined generative model, SAR measurements $\mathbf{y}$ can be simulated using the steering matrix $\mathbf{A}$ of a specific geometric configuration. To ease the network training, we propose to provide as input to the network the \emph{beamforming} profile computed from a multi-look correlation matrix $\mathbf{\widehat{R}}$ obtained from $L$ measurements $\mathbf{y}_1$ to $\mathbf{y}_L$. In this way, the network input and output are in the same space, i.e. both positive-valued vectors, easier to manipulate than complex-valued vectors.
Besides, the \emph{beamforming} profile depends less on the interferometric baselines than the correlation matrix $\mathbf{\widehat{R}}$. It includes the physics of acquisition through the steering matrix.

The correlation matrix has been used rather than the covariance matrix $\boldsymbol{\widehat{\Sigma}}$ to make the method invariant to differing intensity values. The final reconstruction will therefore be $\mathbf{\widehat{p}} = \text{Tr}(\boldsymbol{\widehat{\Sigma}} / N)\,\mathbf{\tilde{p}}$, with $\mathbf{\tilde{p}}$ the output of the network, to recover the actual intensity value of the profile.

We have chosen to use a simple network architecture with an encoder-decoder going through a latent space of reduced dimension. The choice of this dimension is made to represent the number of parameters describing the profile. The loss used is a quadratic loss between the ground-truth simulated profile and the output profile, a standard choice for regression problems. 

\subsection{Generative model}
\label{sec:genemodel}

The distribution of reflectivity in a forest area cell is mainly composed of two peaks, one for the ground and one for the canopy, as validated by the latest tomographic reconstruction methods \cite{Aguilera_2013_WV, Laurent_2022_Dict}. These peaks can be represented by many basis functions, of which the simplest case is that of two Gaussians. The proposed approach therefore aims at learning a latent representation of a forest reflectivity profile from simulated profiles composed of two Gaussians. The parameters of these Gaussians are adapted to the type of forests observed in our SAR datasets for each training, i.e. boreal forests in L band and tropical forests in P band. The ranges of these parameters are chosen to be large, to cover all possible profiles met in practice. The influence of the choice of these parameters is discussed in section \ref{section:discussion}.

\section{Experiments}
\label{section:experiments}

\subsection{Simulated data}

$10000$ simulated profiles sampled on $N_z$ uniformly spaced heights were generated with a mixture of two Gaussians with various parameters. The corresponding covariance matrices $\mathbf{\Sigma}=\mathbf{A} \text{diag}(\mathbf{\mathbf{p}}) \mathbf{A}^H$ were built with a steering matrix $\mathbf{A}$ randomly selected among steering matrices of an actual tomographic dataset. Circular complex Gaussian samples $\{\mathbf{y}_l\}_{l=1..L}$ were then drawn according to these covariance matrices (by multiplying a white speckle noise $\mathbf{w}_l$ by $\mathbf A\text{diag}(\sqrt{\mathbf p})$). Sample correlation matrices $\mathbf{\widehat{R}}$ were computed from each set of $L$ samples $\mathbf{y}_l$.
The \emph{beamforming} profiles were then reconstructed:
\begin{equation}
    {\tilde{{p}}}_i^{\text{(B)}} = \frac{\mathbf{a}^H(z_i) \mathbf{\widehat{R}} \mathbf{a}(z_i)}{N^2}
    \label{eq:bf}
\end{equation}
where $\mathbf{a}$ is a steering column vector of the matrix $\mathbf{A}$. These \emph{beamforming} profiles were then used as input by the neural network.

The trained encoder consists of 4 linear layers with a decreasing number of neurons (starting with $N_z = 512$ and encoding the input data in a latent space of size $5$) and a symmetric decoder of 4 linear layers. The depth of the network is voluntarily kept small and has been empirically validated, as discussed in section \ref{section:architecture}. Each layer is unbiased and has a leaky ReLU activation function. Training was performed on the dataset consisting of the simulated \emph{beamforming} profiles, divided into training (75\%) and validation (25\%) sets, in mini-batches of size 32 with an Adam optimizer and a learning rate of $10^{-3}$ for 200 epochs.

In the following subsections, the results presented were computed using the steering matrices of the BioSAR-2 campaign, more thoroughly presented in section \ref{section:LBand}.

\subsubsection{Analysis of the reconstructed profiles}

\begin{figure}[ht]
    \captionsetup[subfloat]{farskip=2pt,captionskip=0pt,format=hang,margin=10pt,singlelinecheck=false}
    \centering
    \subfloat[Beamforming (blue) \newline Capon (green)]
    {
        \includegraphics[width=0.47\linewidth]{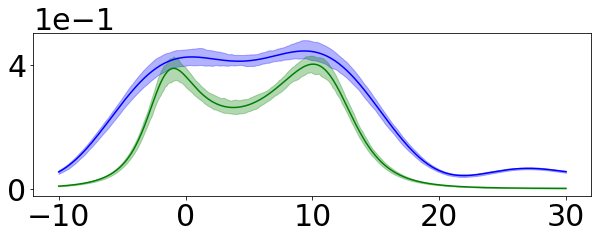}
    }
    \subfloat[Beamforming (blue) \newline Capon (green)]
    {
        \includegraphics[width=0.47\linewidth]{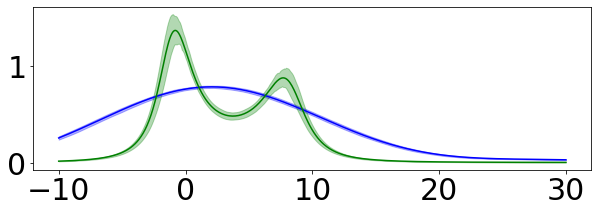}
    }\\
    \subfloat[Ground-truth (black) \newline Proposed method (red) \newline Wavelet-based CS (magenta)]
    {
        \includegraphics[width=0.47\linewidth]{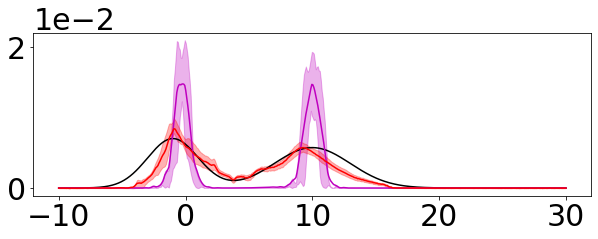}
    }
    \subfloat[Ground-truth (black) \newline Proposed method (red) \newline Wavelet-based CS (magenta)]
    {
        \includegraphics[width=0.47\linewidth]{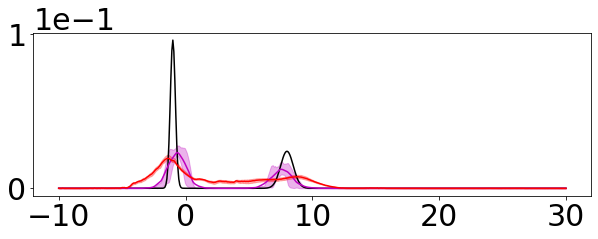}
    }
    \caption{Reconstruction of a profile consisting of wide (left) and narrow (right) Gaussians with Beamforming, Capon, Wavelet-based CS and our method. For each approach, the average profile over 100 measurements with various speckle realizations and its interquartile range have been plotted, with the reference profile shown in black.}
    \label{fig:MSE}
\end{figure}

Fig. \ref{fig:MSE} shows, for two different simulated profiles, the reconstructions obtained with the \emph{beamforming} algorithm, \emph{Capon} filter, \emph{wavelet-based CS} and our method for 100 speckle realizations in the measurement simulation. As we are interested in comparing the shapes of the profiles reconstructed by each algorithm and since they do not compute the same physical quantity, the methods corresponding to the output power of a filter have been plotted above and the reconstructed and reference reflectivity profiles below.
The left profile composed of two large Gaussians favors the spectral estimation methods and shows that the \emph{wavelet-based CS} fails to reconstruct large lobes, while the profile on the right simulates very narrow responses, more suitable for the latter but only roughly reconstructed by \emph{Beamforming} for the level of noise considered ($L = 100$). As for the \emph{Capon} filter, it performs well with low thermal noise, but is very affected when the noise increases, even with suitable diagonal loading. The neural network produces profiles with a width that follows more closely the ground-truth profiles in each case.

\subsubsection{Architecture choice}
\label{section:architecture}

Different latent space sizes were tested to support the intuition of keeping a size corresponding to the number of parameters needed to simulate a mixture of two Gaussians. The number of looks $L$ was set to 100 for this study. The reconstruction errors, compared to a normalized error between the input \emph{beamforming} $\mathbf{\widehat{p}^{\text{(B)}}}$ and reference profile $\mathbf{p}$, $\|\mathbf{\widehat{p}^{\text{(B)}}} \frac{\langle \mathbf{\widehat{p}^{\text{(B)}}} \; , \; \mathbf{p} \rangle}{\langle \mathbf{\widehat{p}^{\text{(B)}}} \; , \; \mathbf{\widehat{p}^{\text{(B)}}} \rangle}-\mathbf{p}\|_2^2$, where $ \langle \cdot \; , \; \cdot \rangle $ is the $L_2$ scalar product, are presented in Table \ref{table:latent_size}. They indicate that a latent space of size smaller than 5 significantly increases the reconstruction error, while a larger latent space brings only marginal improvements.

\begin{table*}
    \centering
    \caption{Mean and standard deviation of the mean squared error (MSE) values after 20 trainings during 200 epochs for each different latent space sizes, compared to a normalized error between the input beamforming and reference profile.}
    \resizebox{\textwidth}{!}{
        \begin{tabular}{l@{\hspace*{2ex}} c@{\hspace*{2ex}} c@{\hspace*{2ex}} c@{\hspace*{2ex}} c@{\hspace*{2ex}} c@{\hspace*{2ex}} c@{\hspace*{2ex}} c@{\hspace*{2ex}} c}
            \toprule
            \textbf{Latent space size} & \textbf{3} & \textbf{4} & \textbf{5} & \textbf{6} & \textbf{8} & \textbf{10} & \textbf{15} & \textbf{20}  \\
            \midrule
    \textbf{MSE ($\times 10^{-1}$)} &  $5.12  \pm  0.10$ &  $4.74  \pm  0.12$ &  $4.65  \pm  0.15$ &  $4.68 \pm 0.19$ &  $4.64 \pm 0.16$ &  $4.68 \pm 0.18$ &  $4.67 \pm 0.15$ &  $4.68 \pm 0.13$  \\
            \bottomrule
        \end{tabular}
    }
    \label{table:latent_size}
\end{table*}

\subsection{Boreal forest at L band}
\label{section:LBand}

\begin{figure}[ht]
    \captionsetup[subfloat]{farskip=2pt,captionskip=0pt}
    \centering
    \subfloat[Beamforming]
    {
        \includegraphics[width=0.97\linewidth]{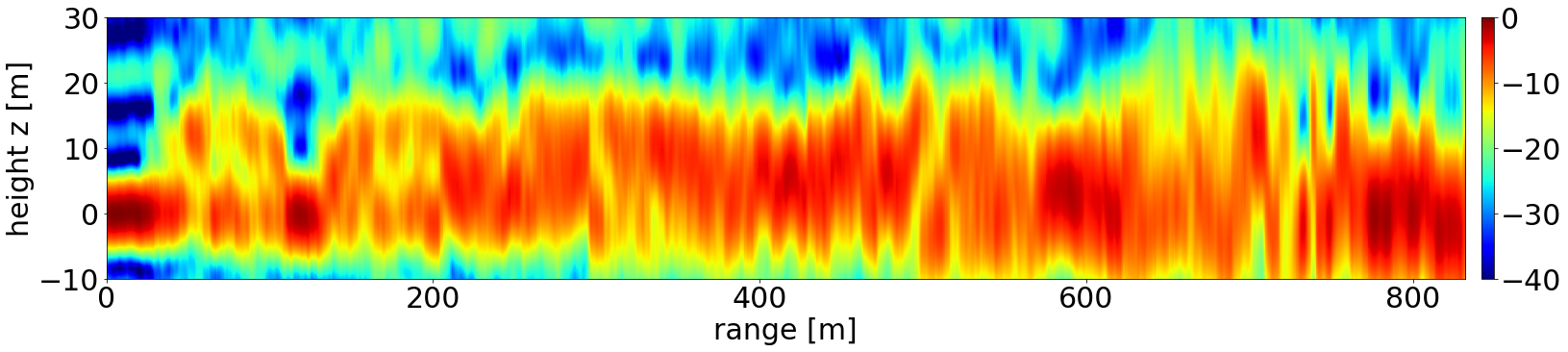}
    }\\
    \subfloat[Capon]
    {
        \includegraphics[width=0.97\linewidth]{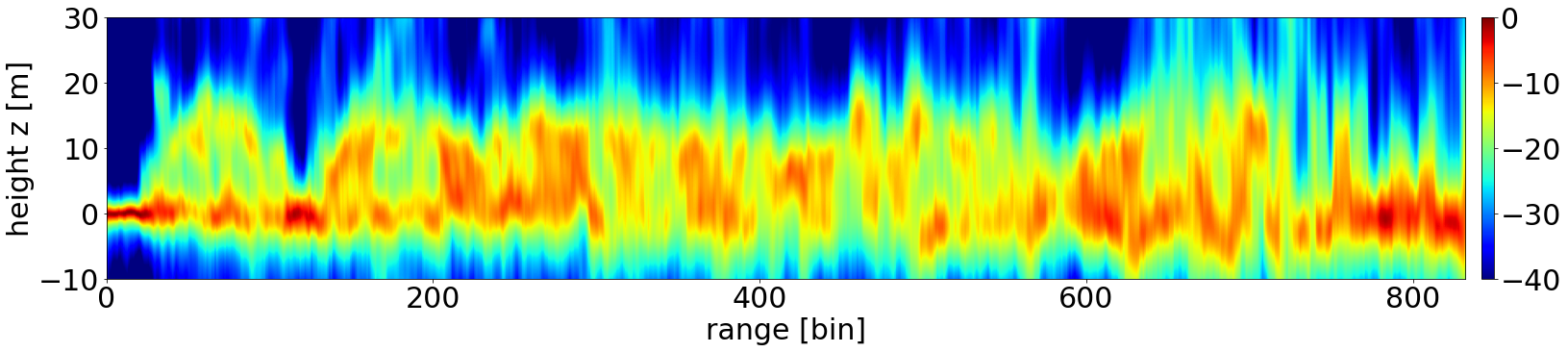}
    }\\
    \subfloat[Wavelet-based CS]
    {
        \includegraphics[width=0.97\linewidth]{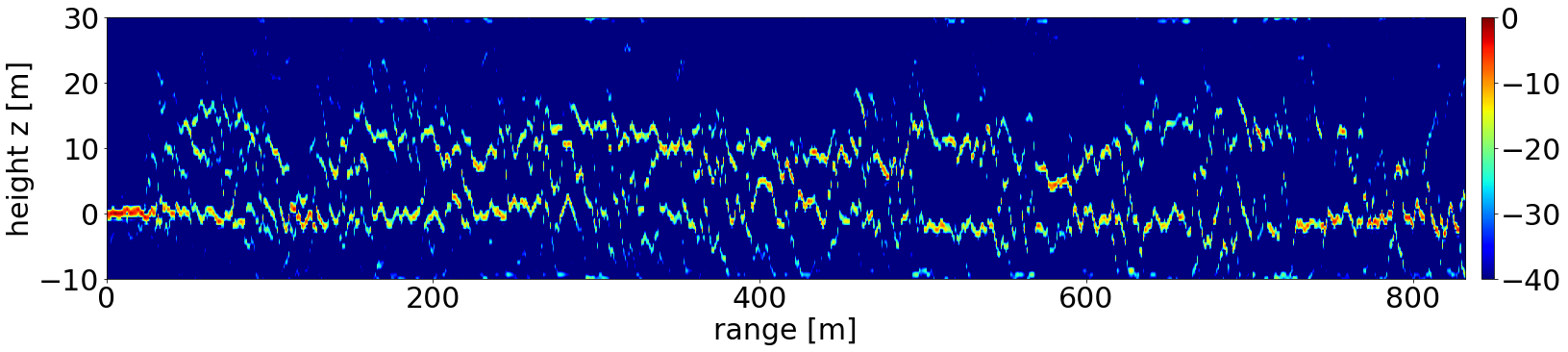}
    }\\
    \subfloat[Proposed method (deep learning)]
    {
        \includegraphics[width=0.97\linewidth]{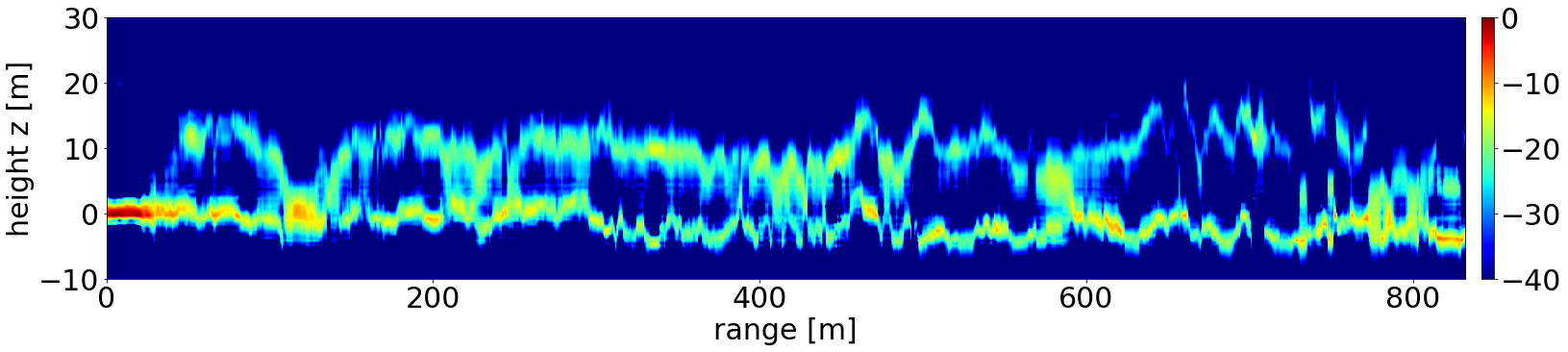}
    }
    \caption{Tomogram for a specific azimuth value in a boreal forest at L band derived with the method: (a) Beamforming; (b) Capon; (c) Wavelet-based CS; (d) Predicted by the proposed neural network.}
    \label{fig:Lband}
\end{figure}

Testing was first performed on a tomographic stack of 6 airborne L-band SAR images of a boreal forest in northern Sweden, acquired during the BioSAR-2 campaign led by the DLR in 2008 \cite{Hajnsek_2008_BioSAR}, with a vertical resolution varying from 6 m in near range to 25 m in far range. A local window of around 60 looks is used to compute the covariance matrix used by the \emph{beamforming} algorithm to compute the tomograms. Fig. \ref{fig:Lband} shows an example of a tomogram in HH polarization reconstructed with the proposed method, after compensation of the topography. It is compared to the results obtained with \emph{beamforming}, \emph{Capon} and the \emph{wavelet-based CS} method developed in \cite{Aguilera_2013_WV}, showing that the trained network does indeed improve the resolution of the tomogram compared to spectral estimation methods, while maintaining a representative volume of the tree crown. This is not the case with the \emph{wavelet-based CS}, which reconstructs narrow peaks and can therefore predict several peaks when this volume is large, depending on the choice of the regularization parameter value.

The orders of magnitude of the computational time required to compute these tomograms are given in Table \ref{table:time} (all computations are done on one CPU, even the network training). The \emph{wavelet-based CS} approach was computed using the optimisation library CVX and the default solver SDPT3 \cite{CVX_2014}. These statistics highlight one of the major advantages of using deep learning to reconstruct forest reflectivity profiles, which is the gain in time of this method compared to classical optimization algorithms. This feature will be even more important when scaling up and applying tomographic reconstruction on an entire SAR image.

\begin{table}[!htpb]
    \centering
    \caption{Order of magnitude of the computation time for the reconstruction of a tomogram of size 1.4 km $\times$ 1.6 m on a cpu.}
    \begin{tabular}{l@{\hspace*{1ex}}l@{\hspace*{1ex}} c}
        \toprule
        \textbf{Method}                             &                       & \textbf{Computation time (s)} \\
        \midrule
        \textbf{Beamforming}                        &                       & 2 \\
        \textbf{Capon}                              &                       & 3 \\
        \textbf{Wavelet-based CS (CVX)}             &                       & 1500 \\
        \multirow{2}{*}{\textbf{Proposed method}$\;\biggl\{$}   & \textbf{Training}     & 200 \\
                                                    & \textbf{Inference}    & 3 \\
        \bottomrule
    \end{tabular}
    \label{table:time}
\end{table}

\subsection{Tropical forest at P band}

\begin{figure}[ht]
    \captionsetup[subfloat]{farskip=2pt,captionskip=0pt}
    \centering
    \subfloat[Beamforming]
    {
        \includegraphics[width=0.97\linewidth]{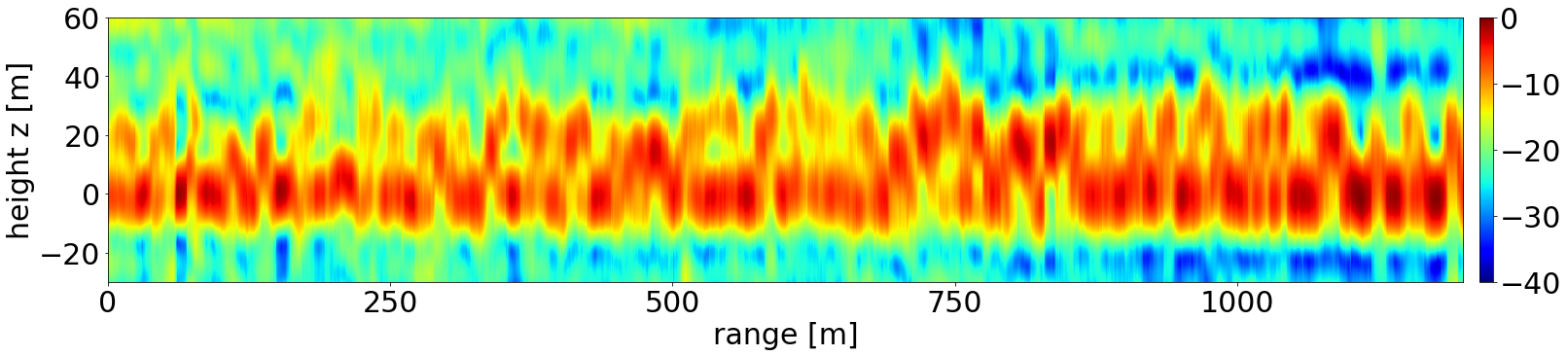}
    }\\
    \subfloat[Capon]
    {
        \includegraphics[width=0.97\linewidth]{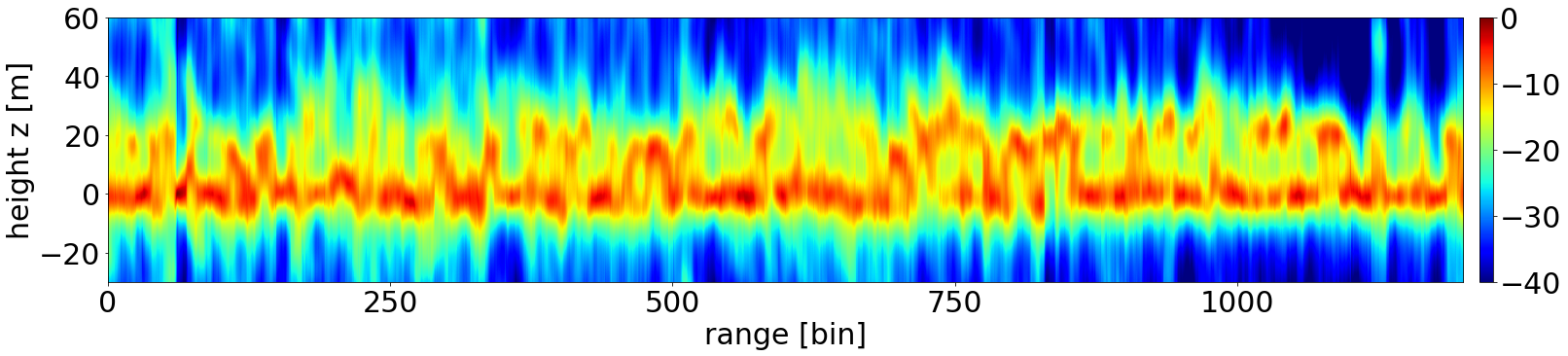}
    }\\
    \subfloat[Wavelet-based CS]
    {
        \includegraphics[width=0.97\linewidth]{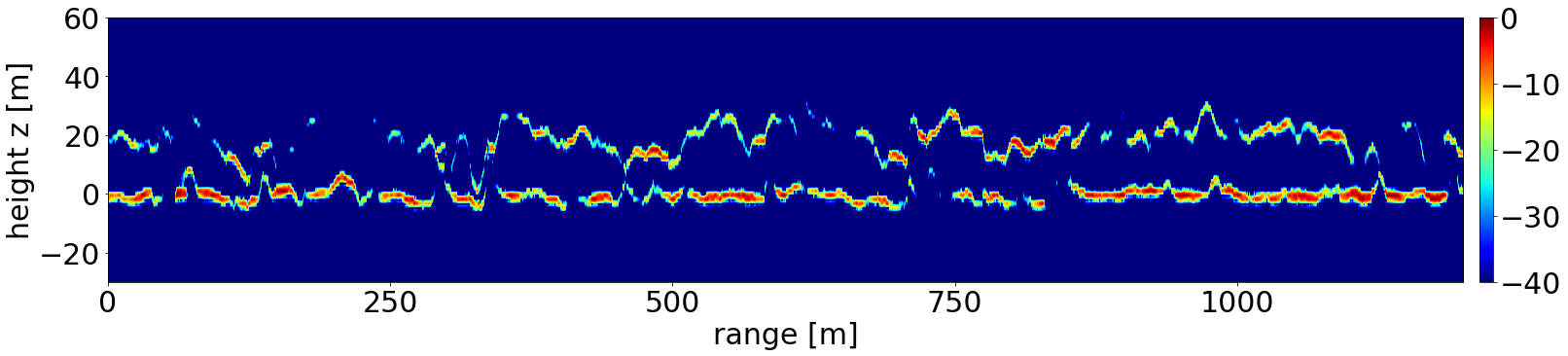}
    }\\
    \subfloat[Proposed method (deep learning)]
    {
        \includegraphics[width=0.97\linewidth]{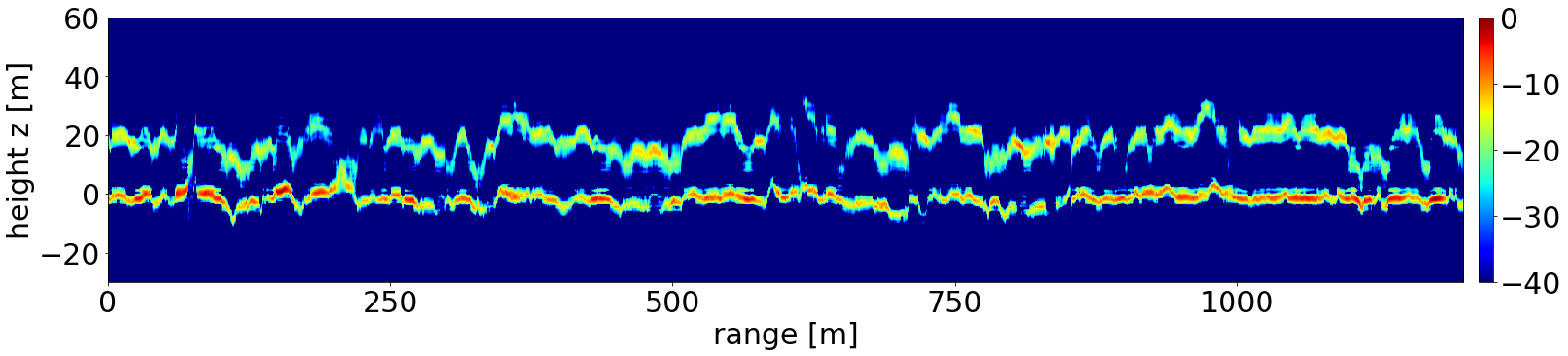}
    }
    \caption{Tomogram for a specific azimuth value in a tropical forest at P band derived with the method: (a) Beamforming; (b) Capon; (c) Wavelet-based CS; (d) Predicted by the proposed neural network.}
    \label{fig:Pband}
\end{figure}

Other tests were performed on airborne tomographic SAR data acquired at P band by the ONERA over the test site of Paracou in French Guiana, during the TropiSAR campaign in 2009 \cite{Dubois_2012_TropiSAR}, also comprising 6 tracks, with a vertical resolution of around 15 m. The correlation matrix used to compare the different methods is calculated with 56 looks. A sample of the results and comparisons are presented in Fig. \ref{fig:Pband}. The proposed approach here seems even more adapted to tropical than boreal forests, providing high resolution reconstructed profiles with fewer information loss than with the \emph{wavelet-based CS}.

\section{Discussion}
\label{section:discussion}

During training, the network is fed with \emph{beamforming} profiles computed from samples generated according to a model of covariance matrix. This model, defined in equation (\ref{eq:cov_forest}), neglects several phenomena such as the integration of back-scattered complex amplitudes produced by scatterers located in the neighborhood according to the SAR impulse response, or the temporal decorrelation between acquisitions. It could be refined, but our results on real data already show a good generalization ability of the network trained with this simple covariance model.

For the training, we used the steering matrix of a specific campaign to simulate measurements and \emph{beamforming} profiles that match at best the actual SAR data considered at test time. The network is thus trained for a given geometric configuration and the efficiency of the network may be affected when it differs, in which case a retraining would be required. However, this step is relatively inexpensive in terms of computational time (a few minutes), and tomographic data acquisition configurations in a given spectral band vary only slightly due to hardware constraints and the possibility of ambiguities.

In our experiments, the network was trained to reconstruct profiles composed of two Gaussians. It cannot therefore perfectly recover a profile with a single or an extra peak, even with input profiles composed of three very distinct Gaussians.
The use of a refined simulation model could further improve the reconstruction quality, which encourages current research on a model representative of different forest types \cite{Laurent_2022_Dict}.

\section{Conclusion}

In this study, we explored the possibility of using deep learning approaches to improve the performance of the tomographic inversion task on forests. Tests on real data show promising results, both in terms of quality and in terms of the computation time needed to reconstruct a large image, with an overhead for the feedforward pass negligible compared to the computation of the \emph{beamforming} profile and an acceleration by several orders of magnitude with respect to iterative regularized inversion algorithms. This will be essential for the systematic processing of future data from the ESA BIOMASS \cite{BIOMASS_report} mission. It confirms the strong potential of the application of deep learning in this field to super-resolve existing reconstructions.

It could also be interesting to consider directly using the measured intensities and acquisition geometry as input to the neural network, to avoid the need to train a different network for each tomographic configuration.


\end{document}